\documentclass[lettersize,journal]{IEEEtran}
\usepackage{cite}

\usepackage{algorithmic}
\usepackage{array}
\usepackage{textcomp}
\usepackage{stfloats}
\usepackage{url}
\usepackage{verbatim}
\usepackage{graphicx}
\usepackage{subfigure}
\usepackage[tbtags]{amsmath}
\usepackage{amsthm,amssymb,bm}
\usepackage{mathrsfs}
\usepackage{enumitem}
\usepackage{tabularx}
\usepackage{threeparttable} 
\usepackage{nicematrix} 
\usepackage{tikz}
\usepackage[colorlinks,bookmarksopen,bookmarksnumbered,citecolor=blue, linkcolor=blue, urlcolor=blue]{hyperref}

\usepackage{booktabs}
\usepackage{multirow}
\usepackage{extarrows}
\usepackage[linesnumbered,ruled,vlined,lined,boxed,commentsnumbered]{algorithm2e}
\usepackage{makecell}
\usepackage{xcolor}
\usepackage{balance}
\def\BibTeX{{\rm B\kern-.05em{\sc i\kern-.025em b}\kern-.08em
		T\kern-.1667em\lower.7ex\hbox{E}\kern-.125emX}}

\SetKwComment{Comment}{/* }{ */}

\begin{document}
	\title{A DSP Framework for ISAC with random signals: From Physical Transceiver to Periodic Models}
	\author{Junjie~Chen 
	\thanks{Manuscript created March 2026; This work was supported by the  (No.xxxxxx). (\textit{Corresponding author: Junjie~Chen.})\\
	Junjie~Chen is with the School of Computing and Artificial Intelligence, Jiangxi University of Finance and Economics, Nanchang 330013, China (e-mail: junjiechen@jxufe.edu.cn).
	}}
	\markboth{IEEE Transactions of xxxx, VOL. xx, NO. xx, NOVEMBER 2026}%
	{How to Use the IEEEtran \LaTeX \ Templates}

	\maketitle

\begin{abstract}
	Communication-centric integrated sensing and communication~(ISAC) directly employs random data-bearing communication signals for both information transmission and environmental sensing. Existing studies generally describe pulse-shaped transmission in continuous time or directly adopt periodic discrete-time sensing models, leaving the digital signal processing~(DSP) operations connecting the physical transceiver and the periodic models unclear. This paper develops a DSP-oriented transceiver framework for single-antenna communication-centric ISAC. A common transmitter is constructed through modulation, cyclic-prefix~(CP) insertion, upsampling, and pulse shaping. For communication reception, the physical transceiver chain is reduced to a symbol-rate equivalent linear channel, which becomes circulant after CP removal and thereby supports frequency-domain equalization. For sensing reception, a high-rate reference waveform is extracted from the transmitted waveform and equivalently represented by circular pulse shaping. CP removal at the receiver then converts physical target delays into circular shifts of this reference waveform, leading to a periodic matched-filtering model and the corresponding range profile. Numerical results validate the developed framework through target range estimation and the symbol error rate performance of pulse-shaped CP-OFDM over frequency-selective Rayleigh fading channels.
\end{abstract}
	
\section{Introduction}

Integrated sensing and communication~(ISAC) is expected to become a fundamental functionality of future wireless networks by enabling sensing and communication~(S\&C) services to share the same spectrum, hardware platform, and transmitted signal~\cite{liu2022integrated}. In general, the signal design methodologies for ISAC can be generally categorized as sensing-centric~\cite{hassanien2015dual}, communication-centric~\cite{xielei2025tcom,shi2026ofdm,xie2026adaptive,he2024dual}, and joint designs~\cite{liu2018toward,liu2021cramer,hua2023mimo}. Among them, communication-centric ISAC is particularly attractive for practical deployment because it directly exploits existing communication signals and infrastructures to support sensing functionalities~\cite{lu2025sensing}. In this architecture, a random data-bearing signal intended for a communication receiver is simultaneously used to probe the surrounding environment, while the corresponding echoes are processed by a sensing receiver to estimate target parameters. Compared with sensing-centric or fully joint waveform designs, this approach preserves the basic communication signal structure and can therefore be integrated into existing wireless systems with relatively minor modifications.

A practical communication waveform is jointly determined by its constellation, modulation basis, and pulse shaping filter~\cite{du2024reshaping,liu2025cpofdm,liu2025uncovering}. The constellation specifies the communication symbols, the modulation basis maps these symbols into discrete-time samples, and the pulse shaping filter generates the band-limited waveform for physical transmission. These components jointly determine the ACF of the transmitted waveform and, consequently, its sensing performance.

Recent studies have made substantial progress in characterizing these waveform-level effects. For communication-centric ISAC signals constructed from arbitrary orthonormal modulation bases, the periodic and aperiodic ACFs were analyzed in~\cite{liu2025cpofdm}. It was proved that, under QAM/PSK constellations, CP-OFDM is the unique globally optimal orthogonal signaling scheme in terms of the expected integrated sidelobe level and each individual sidelobe of the periodic ACF. The analysis was subsequently extended to pulse-shaped random ISAC signals in~\cite{liu2025uncovering}, where a closed-form expression for the average squared ACF was derived under arbitrary modulation bases, constellation mappings, and Nyquist pulse shaping. The resulting ACF was interpreted through the ``iceberg in the sea'' metaphor: the squared mean of the ACF forms the pulse-dependent iceberg, whereas the variance induced by random communication symbols determines the sea level. Related studies have further extended the analysis to the delay-Doppler domain~\cite{zhang2025discrete} and examined the influence of constellation shaping on the communication-sensing tradeoff~\cite{du2024reshaping}.

The signal processing pipeline of communication-centric ISAC was described from a broader information-theoretic and signal-processing perspective in~\cite{liu2025sensing}. In that framework, the pulse-shaped transmit signal, communication multipath channel, communication matched filter, and sensing matched filter are formulated in continuous time through convolution and correlation integrals. Such a representation clearly describes the physical roles of the communication and sensing receivers. Nevertheless, it does not explicitly expose the digital implementation involving CP insertion, upsampling, finite-length pulse shaping, receive filtering, timing alignment, downsampling, and CP removal. Conversely, the discrete-time framework in~\cite{liu2025uncovering} assumes a sufficiently long CP and directly represents the useful pulse-shaped waveform through a circulant pulse-shaping matrix. The resulting sensing echo is then modeled as a superposition of circularly shifted waveform replicas. While this representation is well suited for ACF analysis, the intermediate steps connecting the physical linear waveform generation and propagation processes to the final periodic model are not explicitly presented.

This missing connection leads to several fundamental modeling questions. First, since the CP is inserted into the symbol-rate sequence before upsampling and pulse shaping, it is necessary to clarify how the subsequent physical linear pulse shaping produces a circularly pulse-shaped useful block. Second, at the communication receiver, the CP does not act on the physical multipath channel alone. Instead, it circularizes the symbol-rate equivalent channel formed jointly by upsampling, transmit pulse shaping, physical multipath propagation, receive matched filtering, timing alignment, and downsampling. Third, at the sensing receiver, the physical target channel imposes aperiodic delays on the complete transmitted waveform, whereas periodic ACF analysis requires circular shifts of a fixed-length sensing reference waveform. Therefore, the exact role of CP removal in converting the physical aperiodic target delays into circular shifts needs to be established. Finally, the complete transmitted waveform and the useful sensing reference waveform must be clearly distinguished, since the periodic echo model and the sensing matched filter are expressed in terms of the latter.

Motivated by these observations, this paper develops a unified discrete-time transceiver framework for single-antenna communication-centric ISAC. As illustrated in Fig.~\ref{fig:systemmodel}, the communication and sensing functionalities share a common communication transmitter, while employing separate receivers according to their respective objectives. The transmitter successively performs modulation, CP insertion, upsampling, and pulse shaping to generate the actual waveform entering the physical propagation environment. The communication receiver performs receive matched filtering, timing alignment, downsampling, CP removal, and symbol recovery. In parallel, the sensing receiver extracts a high-rate useful observation block, constructs the corresponding sensing reference waveform, performs periodic matched filtering, and estimates the target ranges from the resulting range profile.

The resulting framework establishes the missing DSP link between physical pulse-shaped CP transmission and the periodic discrete-time models used in random ISAC waveform analysis.
The main contributions of this paper are summarized as follows:
\begin{itemize}
	\item A complete discrete-time model is established for the common ISAC transmitter. In contrast to models that directly assume circular pulse shaping, the proposed formulation explicitly describes the sequence of modulation, CP insertion, upsampling, and finite-length linear pulse shaping, thereby distinguishing the complete transmitted waveform from the high-rate useful sensing reference waveform.
	\item A unified communication receiver model is derived by combining the transmit pulse, physical multipath channel, receive matched filter, timing alignment, and downsampling into a symbol-rate equivalent linear channel. Under an explicit CP-length condition, CP insertion and removal convert this equivalent linear channel into a circulant channel, which can then be diagonalized and equalized in the frequency domain for OFDM signaling.
	\item A discrete-time sensing receiver model is developed from the physical aperiodic target-echo model. The useful sensing reference waveform is shown to admit two equivalent representations: one obtained by extracting the useful interval from the actual transmitted waveform, and the other obtained through circular pulse shaping of the CP-free upsampled signal. Under a sufficiently long CP, the physical aperiodic target delays are converted into circular shifts of this reference waveform, yielding the periodic matched-filtering and range-profile models adopted in random ISAC signal analysis.
	\item The resulting framework establishes an explicit mathematical connection between the continuous-time transceiver model in~\cite{liu2025sensing} and the periodic discrete-time waveform model in~\cite{liu2025uncovering}. Numerical results validate the derived models through periodic matched-filter-based range estimation and the SER performance of pulse-shaped CP-OFDM over frequency-selective Rayleigh fading channels.
\end{itemize}

The remainder of this paper is organized as follows. 
Section~II describes the physical transmission and sensing processes and derives the corresponding signal models with CP insertion and removal.
Section~III summarizes the equivalent end-to-end communication and sensing models induced by CP insertion and removal. 
Section~IV verifies the derived operator identities and illustrates the resulting communication symbol recovery and sensing range profiles. 
Finally, Section~V concludes the paper and discusses possible extensions.

\section{Physical Transceiver and CP-Induced Signal Models}

\begin{figure*}[t] 
	\centering
	\includegraphics[width=1\textwidth]{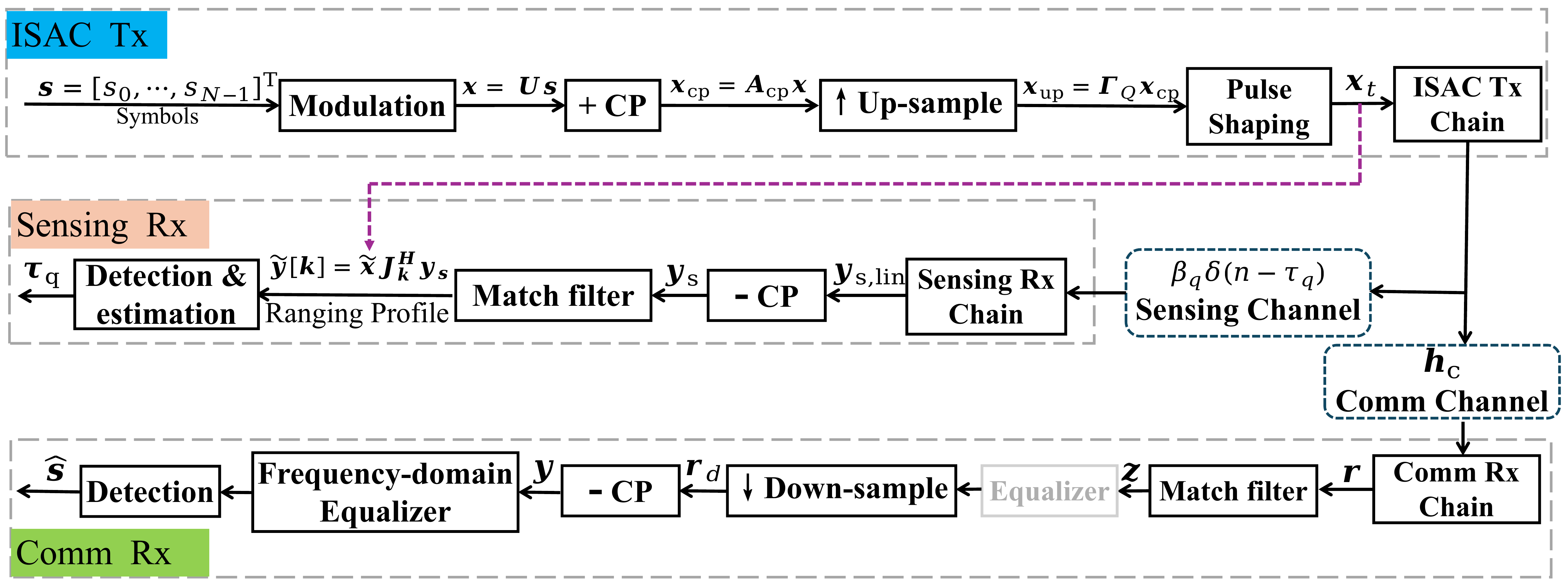}
	\caption{The equivalent DSP implementation of a communication-centric ISAC system.}
	\label{fig:systemmodel}
\end{figure*}

Fig.~\ref{fig:systemmodel} illustrates the complete discrete-time signal processing architecture considered in this paper. The system consists of a common single-antenna ISAC transmitter, a communication receiver, and a sensing receiver. The common transmitter follows the standard communication waveform generation procedure.

\subsection{ISAC Transmitter}

As illustrated in Fig.~\ref{fig:systemmodel}, we consider a single-antenna communication-centric ISAC system, where the communication and sensing functionalities share a common transmitter.

\subsubsection{Modulation}

Let $\bm{s}=[s[0],s[1],\ldots,s[N-1]]^{\mathrm T}\in\mathbb C^{N\times1}$ denote a block of $N$ communication symbols. These symbols are modulated over an orthonormal basis represented by a unitary matrix $\bm{U}=[\bm{u}_0,\bm{u}_1,\ldots,\bm{u}_{N-1}]\in\mathbb U(N)$, where $\mathbb U(N)$ denotes the unitary group of degree $N$. The resulting discrete-time signal is expressed as
\begin{equation}
	\begin{aligned}
		\bm{x}=\bm{U}\bm{s}=\sum_{n=0}^{N-1}s[n]\bm{u}_n\in\mathbb C^{N\times1}.
	\end{aligned}
	\label{eq:x_modulated}
\end{equation}
The above formulation accommodates various communication signaling schemes through different choices of $\bm{U}$. Some representative examples are given as follows:
\begin{itemize}
	\item \textbf{SC:} $\bm{U}=\bm{I}_N$. In this case, the communication symbols are transmitted consecutively in the time domain, and the resulting $\bm{x}$ is a single-carrier signal.
	\item \textbf{OFDM:} $\bm{U}=\bm{F}_N^{\mathrm H}$, where $\bm{F}_N$ is the normalized $N$-point DFT matrix. The communication symbols are placed in the frequency domain, and the resulting $\bm{x}$ is an OFDM signal with $N$ subcarriers.
	\item \textbf{CDMA:} $\bm{U}=\bm{C}_N$, where $\bm{C}_N$ is the normalized Hadamard matrix of size $N$. The communication symbols are placed in the code domain, and the resulting $\bm{x}$ is a CDMA signal.
	\item \textbf{OTFS:} $\bm{U}=\bm{F}_{N_{\mathrm t}}^{\mathrm H}\otimes\bm{I}_{N_{\mathrm f}}$, where $N=N_{\mathrm t}N_{\mathrm f}$, and $N_{\mathrm t}$ and $N_{\mathrm f}$ denote the numbers of time slots and subcarriers, respectively. In this case, $\bm{s}$ is placed in the delay-Doppler domain.
	\item \textbf{AFDM:} $\bm{U}=\bm{\Lambda}_{c_1}^{\mathrm H}\bm{F}_N^{\mathrm H}\bm{\Lambda}_{c_2}^{\mathrm H}$, where
	\begin{equation}
		\begin{aligned}
			\bm{\Lambda}_{c}=\operatorname{Diag}\left(1,e^{-j2\pi c1^2},\ldots,e^{-j2\pi c(N-1)^2}\right).
		\end{aligned}
	\end{equation}
	In this case, $\bm{U}$ is the inverse discrete affine Fourier transform matrix, and the communication symbols are placed in the affine Fourier transform domain.
\end{itemize}

\subsubsection{CP Insertion}

To facilitate efficient signal processing in the frequency domain at the communication receiver and the periodic correction at the sensing receiver, a cyclic prefix (CP) of length $N_{\mathrm{cp}}$ is added to $\bm{x}$. Let $M=N+N_{\mathrm{cp}}$. The CP-extended signal is expressed as
\begin{equation}
	\begin{aligned}
		\bm{x}_{\mathrm{cp}} =\bm{A}_{\mathrm{cp}}\bm{x} =\bm{A}_{\mathrm{cp}}\bm{U}\bm{s}\in\mathbb C^{M\times1},
	\end{aligned}
	\label{eq:x_cp}
\end{equation}
where the CP-insertion matrix is given by
\begin{equation}
	\begin{aligned}
		\bm{A}_{\mathrm{cp}}=\begin{bmatrix}\bm{0}_{N_{\mathrm{cp}}\times(N-N_{\mathrm{cp}})}&\bm{I}_{N_{\mathrm{cp}}}\\\bm{I}_N\end{bmatrix}\in\mathbb R^{M\times N}.
	\end{aligned}
	\label{eq:A_cp}
\end{equation}

\subsubsection{Upsampling}

The CP-extended signal is subsequently upsampled by a over-sampling ratio~$Q$. Define the upsampling matrix $\bm{\Gamma}_Q\in\mathbb R^{QM\times M}$ by
\begin{equation}
	\begin{aligned}
		[\bm{\Gamma}_Q]_{n,m}=\begin{cases}1,&n=mQ,\\0,&\text{otherwise},\end{cases}
	\end{aligned}
	\label{eq:Gamma_Q}
\end{equation}
where $0\leq n\leq QM-1$ and $0\leq m\leq M-1$. The upsampled signal is therefore given by
\begin{equation}
	\begin{aligned}
		\bm{x}_{\mathrm{up}}=\bm{\Gamma}_Q\bm{x}_{\mathrm{cp}}\in\mathbb C^{QM\times1}.
	\end{aligned}
	\label{eq:x_up}
\end{equation}

\subsubsection{Pulse Shaping}

For subsequent use, we first define the full linear convolution matrix. Consider a finite impulse response $\bm{a}=[a[0],a[1],\ldots,a[L_a-1]]^{\mathrm T}\in\mathbb C^{L_a\times1}$. For an input vector of length $K$, the full linear convolution matrix generated by $\bm{a}$ is defined as
\begin{equation}
	\begin{aligned}
		&\bm{T}(\bm{a};K)= \\
		&~\begin{bmatrix}a[0]&0&\cdots&0\\a[1]&a[0]&\ddots&\vdots\\\vdots&a[1]&\ddots&0\\a[L_a-1]&\vdots&\ddots&a[0]\\0&a[L_a-1]&\ddots&a[1]\\\vdots&\ddots&\ddots&\vdots\\0&\cdots&0&a[L_a-1]\end{bmatrix}\in\mathbb C^{(K+L_a-1)\times K}.
	\end{aligned}
	\label{eq:linear_convolution_matrix}
\end{equation}
For any $\bm{v}\in\mathbb C^{K\times1}$, $\bm{T}(\bm{a};K)\bm{v}$ gives the full linear convolution between $\bm{a}$ and $\bm{v}$.

Let $\bm{p}_{\mathrm t}=[p_{\mathrm t}[0],p_{\mathrm t}[1],\ldots,p_{\mathrm t}[L_{\mathrm p}-1]]^{\mathrm T}\in\mathbb C^{L_{\mathrm p}\times1}$ denote the transmit pulse. The corresponding pulse-shaping matrix is
\begin{equation}
	\begin{aligned}
		\bm{P}_{\mathrm t}\triangleq\bm{T}(\bm{p}_{\mathrm t};QM)\in\mathbb C^{(QM+L_{\mathrm p}-1)\times QM}.
	\end{aligned}
	\label{eq:P_t}
\end{equation}
After pulse shaping, the actual waveform radiated by the common ISAC transmitter is expressed as
\begin{equation}
	\begin{aligned}
		\boxed{\bm{x}_{\mathrm t}=\bm{P}_{\mathrm t}\bm{\Gamma}_Q\bm{A}_{\mathrm{cp}}\bm{U}\bm{s}\in\mathbb C^{K_{\mathrm t}\times1}},
	\end{aligned}
	\label{eq:x_transmit}
\end{equation}
where $K_{\mathrm t}=QM+L_{\mathrm p}-1$. Therefore, the communication and sensing receivers shown in Fig.~\ref{fig:systemmodel} process different observations of the same transmitted waveform $\bm{x}_{\mathrm t}$.

\subsection{Communication Receiver}

\subsubsection{Communication Channel}

Let $\bm{h}=[h[0],h[1],\ldots,h[L-1]]^{\mathrm T}\in\mathbb C^{L\times1}$ denote the impulse response of the frequency-selective communication channel. The corresponding full linear convolution matrix is
\begin{equation}
	\begin{aligned}
		\bm{H}\triangleq\bm{T}(\bm{h};QM+L_{\mathrm p}-1)\in\mathbb C^{(QM+L_{\mathrm p}+L-2)\times(QM+L_{\mathrm p}-1)}.
	\end{aligned}
	\label{eq:H_comm}
\end{equation}
The received high-rate communication signal is given by
\begin{equation}
	\begin{aligned}
		\bm{r}=\bm{H}\bm{x}_{\mathrm t}+\bm{w}=\bm{H}\bm{P}_{\mathrm t}\bm{\Gamma}_Q\bm{A}_{\mathrm{cp}}\bm{U}\bm{s}+\bm{w},
	\end{aligned}
	\label{eq:comm_received}
\end{equation}
where $\bm{w}$ denotes the additive noise. The multiplication by $\bm{H}$ represents the physical linear convolution between the transmitted waveform and the multipath communication channel.

\subsubsection{Matched Filtering}

The communication receiver first applies a receive filter consisting of the matched filter and the optional time-domain equalizer shown in gray in Fig.~\ref{fig:systemmodel}. In practical systems, such an equalizer is used to mitigate the residual ISI caused by nonideal pulse shaping and the physical channel. Since the cascaded receive filtering and equalization operations are linear, their combined response is denoted by $p_{\mathrm r}[n]$ in the following derivation. When no separate time-domain equalizer is employed, $p_{\mathrm r}[n]$ reduces to the conventional matched-filter response
\begin{equation}
	\begin{aligned}
		p_{\mathrm r}[n]=p_{\mathrm t}^{*}[L_{\mathrm p}-1-n].
	\end{aligned}
	\label{eq:matched_filter}
\end{equation}
Let $\bm{p}_{\mathrm r}=[p_{\mathrm r}[0],p_{\mathrm r}[1],\ldots,p_{\mathrm r}[L_{\mathrm r}-1]]^{\mathrm T}\in\mathbb C^{L_{\mathrm r}\times1}$ and define
\begin{equation}
	\begin{aligned}
		\bm{P}_{\mathrm r}\triangleq\bm{T}(\bm{p}_{\mathrm r};QM+L_{\mathrm p}+L-2).
	\end{aligned}
	\label{eq:P_r}
\end{equation}
The matched-filter output is expressed as
\begin{equation}
	\begin{aligned}
		\bm{z}=\bm{P}_{\mathrm r}\bm{H}\bm{P}_{\mathrm t}\bm{\Gamma}_Q\bm{A}_{\mathrm{cp}}\bm{U}\bm{s}+\widetilde{\bm{w}},
	\end{aligned}
	\label{eq:z_comm}
\end{equation}
where $\widetilde{\bm{w}}=\bm{P}_{\mathrm r}\bm{w}$.

The transmit pulse, physical communication channel, and receive matched filter can be combined into the high-rate equivalent impulse response
\begin{equation}
	\begin{aligned}
		c[n]\triangleq\left(p_{\mathrm r}*h*p_{\mathrm t}\right)[n].
	\end{aligned}
	\label{eq:c_equivalent}
\end{equation}
The length of $c[n]$ is
\begin{equation}
	\begin{aligned}
		L_c=L_{\mathrm p}+L+L_{\mathrm r}-2.
	\end{aligned}
	\label{eq:Lc}
\end{equation}
Define $\bm{c}=[c[0],c[1],\ldots,c[L_c-1]]^{\mathrm T}\in\mathbb C^{L_c\times1}$. The linear convolution matrix generated by $\bm{c}$ is
\begin{equation}
	\begin{aligned}
		&\bm{C}\triangleq\bm{T}(\bm{c};QM)=\\
		&~\begin{bmatrix}c[0]&0&\cdots&0\\c[1]&c[0]&\ddots&\vdots\\\vdots&c[1]&\ddots&0\\c[L_c-1]&\vdots&\ddots&c[0]\\0&c[L_c-1]&\ddots&c[1]\\\vdots&\ddots&\ddots&\vdots\\0&\cdots&0&c[L_c-1]\end{bmatrix}\in\mathbb C^{(QM+L_c-1)\times QM}.
	\end{aligned}
	\label{eq:C_matrix}
\end{equation}
By the associativity of linear convolution,
\begin{equation}
	\begin{aligned}
		\bm{P}_{\mathrm r}\bm{H}\bm{P}_{\mathrm t}=\bm{C}=\bm{T}(\bm{c};QM).
	\end{aligned}
	\label{eq:convolution_associativity}
\end{equation}
Accordingly, the matched-filter output can be equivalently written as
\begin{equation}
	\begin{aligned}
		\bm{z}=\bm{T}(\bm{c};QM)\bm{\Gamma}_Q\bm{x}_{\mathrm{cp}}+\widetilde{\bm{w}}.
	\end{aligned}
	\label{eq:z_equivalent}
\end{equation}

\subsubsection{Downsampling}

By the definition of linear convolution, the $n$th sample of \eqref{eq:z_equivalent} is
\begin{equation}
	\begin{aligned}
		z[n]=\sum_{a=0}^{QM-1}x_{\mathrm{up}}[a]c[n-a]+\widetilde{w}[n].
	\end{aligned}
	\label{eq:z_sample}
\end{equation}
Since the upsampled sequence satisfies $x_{\mathrm{up}}[jQ]=x_{\mathrm{cp}}[j]$ and is zero at the remaining high-rate sampling instants,
\begin{equation}
	\begin{aligned}
		z[n]=\sum_{j=0}^{M-1}x_{\mathrm{cp}}[j]c[n-jQ]+\widetilde{w}[n].
	\end{aligned}
	\label{eq:z_sample_cp}
\end{equation}
Let $n_0$ denote the correct sampling phase. Starting from $n_0$, the receiver retains one sample every $Q$ high-rate samples, yielding
\begin{equation}
	\begin{aligned}
		r_{\mathrm d}[m]&=z[n_0+mQ]\\&=\sum_{j=0}^{M-1}x_{\mathrm{cp}}[j]c[n_0+(m-j)Q]+w_{\mathrm d}[m],
	\end{aligned}
	\label{eq:r_downsampled}
\end{equation}
where $w_{\mathrm d}[m]=\widetilde{w}[n_0+mQ]$.

Define the symbol-rate equivalent channel as
\begin{equation}
	\begin{aligned}
		h_{\mathrm{eq}}[k]\triangleq c[n_0+kQ],\qquad k=0,1,\ldots,L_{\mathrm{eq}}-1,
	\end{aligned}
	\label{eq:h_eq}
\end{equation}
where
\begin{equation}
	\begin{aligned}
		L_{\mathrm{eq}}=\left\lfloor\frac{L_c-1-n_0}{Q}\right\rfloor+1.
	\end{aligned} \label{eq:L_eq}
\end{equation}
The corresponding equivalent channel vector is
\begin{equation}
	\begin{aligned}
		\bm{h}_{\mathrm{eq}}=[c[n_0],c[n_0+Q],\ldots,c[n_0+(L_{\mathrm{eq}}-1)Q]]^{\mathrm T}\in\mathbb C^{L_{\mathrm{eq}}\times1}.
	\end{aligned}
	\label{eq:h_eq_vector}
\end{equation}
Consequently,
\begin{equation}
	\begin{aligned}
		r_{\mathrm d}[m]=\sum_{j=0}^{M-1}x_{\mathrm{cp}}[j]h_{\mathrm{eq}}[m-j]+w_{\mathrm d}[m].
	\end{aligned} \label{eq:r_d_convolution}
\end{equation}

Define the downsampling matrix $\bm{D}_{Q,n_0}\in\mathbb R^{(M+L_{\mathrm{eq}}-1)\times(QM+L_c-1)}$ whose elements are
\begin{equation}
	\begin{aligned}
		[\bm{D}_{Q,n_0}]_{m,n}=\begin{cases}1,&n=n_0+mQ,\\0,&\text{otherwise}.\end{cases}
	\end{aligned} \label{eq:D_Q}
\end{equation}
The downsampled received vector is
\begin{equation}
	\begin{aligned}
		\bm{r}_{\mathrm d}=\bm{D}_{Q,n_0}\bm{z}.
	\end{aligned} \label{eq:r_d_downsampling}
\end{equation}
Define the symbol-rate equivalent linear channel matrix as
\begin{equation}
	\begin{aligned}
		\bm{H}_{\mathrm{eq,lin}}\triangleq\bm{T}(\bm{h}_{\mathrm{eq}};M)\in\mathbb C^{(M+L_{\mathrm{eq}}-1)\times M}.
	\end{aligned} \label{eq:H_eq_lin}
\end{equation}
Since $h_{\mathrm{eq}}[k]=c[n_0+kQ]$, it follows that
\begin{equation}
	\begin{aligned}
		\boxed{\bm{H}_{\mathrm{eq,lin}}=\bm{D}_{Q,n_0}\bm{T}(\bm{c};QM)\bm{\Gamma}_Q}.
	\end{aligned} \label{eq:H_eq_identity}
\end{equation}
Therefore, the cascade of upsampling, transmit pulse shaping, the physical communication channel, receive matched filtering, timing alignment, and downsampling is equivalent to the symbol-rate linear channel $\bm{H}_{\mathrm{eq,lin}}$. The downsampled signal is ultimate expressed as
\begin{equation}
	\begin{aligned}
	 \bm{r}_{\mathrm d}=\bm{H}_{\mathrm{eq,lin}}\bm{x}_{\mathrm{cp}}+\bm{w}_{\mathrm d} ,
	\end{aligned} \label{eq:r_d_matrix}
\end{equation}
where $\bm{w}_{\mathrm d}=\bm{D}_{Q,n_0}\widetilde{\bm{w}}$.

\subsubsection{CP Removal}

The communication receiver removes the CP and retains the subsequent $N$ symbol-rate samples. Define
\begin{equation}
	\begin{aligned}
		\overline{\bm{R}}_{\mathrm{cp}}=\begin{bmatrix}\bm{0}_{N\times N_{\mathrm{cp}}}&\bm{I}_N&\bm{0}_{N\times(L_{\mathrm{eq}}-1)}\end{bmatrix}\in\mathbb R^{N\times(M+L_{\mathrm{eq}}-1)}.
	\end{aligned} \label{eq:R_cp_bar}
\end{equation}
The useful communication block is given by
\begin{equation}
	\begin{aligned}
		\bm{y}=\overline{\bm{R}}_{\mathrm{cp}}\bm{r}_{\mathrm d}.
	\end{aligned} \label{eq:y_comm}
\end{equation}
Assume that the CP length satisfies
\begin{equation}
	\begin{aligned}
		N_{\mathrm{cp}}\geq L_{\mathrm{eq}}-1,
	\end{aligned} \nonumber \label{eq:comm_cp_condition}
\end{equation}
the CP insertion and removal operations convert the symbol-rate equivalent linear channel into a circulant channel~\cite{tse2005fundamentals}:
\begin{equation}
	\begin{aligned}
	 \overline{\bm{R}}_{\mathrm{cp}}\bm{H}_{\mathrm{eq,lin}}\bm{A}_{\mathrm{cp}}=\bm{H}_{\mathrm{eq,cir}} ,
	\end{aligned} \label{eq:comm_circularization}
\end{equation}
where $\bm{H}_{\mathrm{eq,cir}}\in\mathbb C^{N\times N}$ is the circulant matrix generated by $\bm{h}_{\mathrm{eq}}$. Therefore,
\begin{equation}
	\begin{aligned}
		\boxed{\bm{y}=\bm{H}_{\mathrm{eq,cir}}\bm{U}\bm{s}+\bm{w}_y.}
	\end{aligned} \label{eq:comm_useful_block}
\end{equation}
Thus, the CP circularizes the symbol-rate equivalent channel formed by the complete pulse-shaped communication link, rather than the physical channel $\bm{h}$ alone.

\subsubsection{Symbol Recovery}
For CP-based block transmission, the frequency-domain equalizer shown in Fig.~\ref{fig:systemmodel} provides a low-complexity alternative to long time-domain equalizers after the equivalent channel has been circularized by CP insertion and removal.
For OFDM signaling, $\bm{U}=\bm{F}_N^{\mathrm H}$. Since $\bm{H}_{\mathrm{eq,cir}}$ is circulant, it can be diagonalized by the normalized DFT matrix as
\begin{equation}
	\begin{aligned}
		\bm{H}_{\mathrm{eq,cir}}=\bm{F}_N^{\mathrm H}\bm{\Lambda}_{\mathrm{eq}}\bm{F}_N,
	\end{aligned} \label{eq:H_eq_diagonalization}
\end{equation}
where~$\bm{\Lambda}_{\mathrm{eq}}=\sqrt{N}\operatorname{Diag}\left(\bm{F}_N\bm{h}_{\mathrm{eq},N}\right)$ and $\bm{h}_{\mathrm{eq},N}$ is obtained by zero-padding $\bm{h}_{\mathrm{eq}}$ to length $N$. Applying the $N$-point DFT to $\bm{y}$ yields
\begin{equation}
	\begin{aligned}
		\bm{Y}=\bm{F}_N\bm{y}=\bm{\Lambda}_{\mathrm{eq}}\bm{s}+\bm{W}.
	\end{aligned} \label{eq:Y_comm}
\end{equation}
With perfect knowledge of the equivalent channel, zero-forcing equalization gives
\begin{equation}
	\begin{aligned}
		\widehat{\bm{s}}=\bm{\Lambda}_{\mathrm{eq}}^{-1}\bm{F}_N\bm{y}.
	\end{aligned} \label{eq:s_recovered}
\end{equation}
In the absence of noise and provided that all diagonal entries of $\bm{\Lambda}_{\mathrm{eq}}$ are nonzero, $\widehat{\bm{s}}=\bm{s}$.

\subsection{Sensing Receiver}

\subsubsection{Sensing Channel}

Different from the communication receiver, the sensing receiver observes delayed and scaled replicas of the common transmitted waveform $\bm{x}_{\mathrm t}$. We consider integer sample delays and ignore Doppler shifts in the following derivation. Suppose that the sensing scene contains $K$ targets. Let $\beta_q\in\mathbb C$ and $\tau_q\in\mathbb Z_{\geq0}$ denote the complex reflection coefficient and high-rate delay index of the $q$th target, respectively. For a high-rate sampling frequency $F_{\mathrm s}=1/T_{\mathrm s}$, the corresponding target range is
\begin{equation}
	\begin{aligned}
		R_q =\frac{c_0T_{\mathrm s}}{2}\tau_q =\frac{c_0}{2F_{\mathrm s}}\tau_q,
	\end{aligned} 	\label{eq:range_delay_mapping}
\end{equation}
where $c_0$ denotes the speed of light. Accordingly, one high-rate delay sample corresponds to
\begin{equation}
	\begin{aligned}
		\Delta R =\frac{c_0T_{\mathrm s}}{2} =\frac{c_0}{2F_{\mathrm s}}.
	\end{aligned} \nonumber \label{eq:range_resolution_grid}
\end{equation}

Let $\tau_{\max}=\max_{1\leq q\leq K}\tau_q$. Since $\bm{x}_{\mathrm t}$ has length $K_{\mathrm t}$, the complete sensing observation has length $K_{\mathrm s}=K_{\mathrm t}+\tau_{\max}$. The physical delay associated with the $q$th target is represented by the aperiodic delay matrix $\widetilde{\bm{J}}_{\tau_q}\in\mathbb R^{K_{\mathrm s}\times K_{\mathrm t}}$, whose elements are
\begin{equation}
	\begin{aligned}
		[\widetilde{\bm{J}}_{\tau_q}]_{n,m}=\begin{cases}1,&n=m+\tau_q,\\0,&\text{otherwise}.\end{cases}
	\end{aligned} \label{eq:J_tilde}
\end{equation}
Ignoring Doppler shifts, the complete sensing echo is expressed as
\begin{equation}
	\begin{aligned}
		\boxed{\bm{y}_{\mathrm{s,lin}}=\sum_{q=1}^{K}\beta_q\widetilde{\bm{J}}_{\tau_q}\bm{x}_{\mathrm t}+\bm{z}_{\mathrm{s,lin}}},
	\end{aligned} \label{eq:sensing_full_echo}
\end{equation}
where $\bm{z}_{\mathrm{s,full}}$ denotes the sensing receiver noise. This model describes the physical aperiodic delays of the transmitted waveform.

The sensing processor constructs a high-rate reference waveform from the known transmitted symbols and transmit processing chain. Define
\begin{equation}
	\begin{aligned}
		\bm{R}_{\mathrm t}=\begin{bmatrix}\bm{0}_{QN\times QN_{\mathrm{cp}}}&\bm{I}_{QN}&\bm{0}_{QN\times(L_{\mathrm p}-1)}\end{bmatrix}\in\mathbb R^{QN\times K_{\mathrm t}}.
	\end{aligned} \label{eq:R_t}
\end{equation}
The first representation of the sensing reference waveform is
\begin{equation}
	\begin{aligned}
		\boxed{\widetilde{\bm{x}}=\bm{R}_{\mathrm t}\bm{x}_{\mathrm t}=\bm{R}_{\mathrm t}\bm{P}_{\mathrm t}\bm{\Gamma}_Q\bm{A}_{\mathrm{cp}}\bm{U}\bm{s}\in\mathbb C^{QN\times1}}.
	\end{aligned} \label{eq:x_tilde}
\end{equation}
Here, $\bm{R}_{\mathrm t}$ is a local reference-selection operation at the sensing processor and does not alter the waveform physically transmitted by the ISAC transmitter.

To obtain an equivalent periodic representation of $\widetilde{\bm{x}}$, define the $Q$-fold upsampling matrix without CP as $\bm{\Gamma}_{Q,N}\in\mathbb R^{QN\times N}$, whose elements satisfy
\begin{equation}
	\begin{aligned}
		[\bm{\Gamma}_{Q,N}]_{n,m}=\begin{cases}1,&n=mQ,\\0,&\text{otherwise}.\end{cases}
	\end{aligned} \label{eq:Gamma_QN}
\end{equation}
Define the length-$QN$ zero-padded transmit pulse as
\begin{equation}
	\begin{aligned}
		\bm{p}_{\mathrm{t},QN}=\begin{bmatrix}\bm{p}_{\mathrm t}\\\bm{0}_{QN-L_{\mathrm p}}\end{bmatrix}\in\mathbb C^{QN\times1}
	\end{aligned} \label{eq:p_t_QN}
\end{equation}
and the corresponding circular pulse-shaping matrix as
\begin{equation}
	\begin{aligned}
		\bm{P}_{\mathrm{t,cir}}=\operatorname{Circ}(\bm{p}_{\mathrm{t},QN})\in\mathbb C^{QN\times QN}.
	\end{aligned} \label{eq:P_t_cir}
\end{equation}
Here, $\operatorname{Circ}(\bm{p}_{\mathrm{t},QN})\in\mathbb C^{QN\times QN}$ denotes the circulant matrix generated by $\bm{p}_{\mathrm{t},QN}$, with $\bm{p}_{\mathrm{t},QN}$ as its first column. For any $\bm{a},\bm{b}\in\mathbb C^{QN\times1}$, $\operatorname{Circ}(\bm{a})\bm{b}=\bm{a}\circledast\bm{b}$, where $\circledast$ denotes circular convolution between two length-$QN$ vectors. 
Assuming that the high-rate CP covers the memory of the transmit pulse, i.e.,~$QN_{\mathrm{cp}}\geq L_{\mathrm p}-1$,
then following equality holds
\begin{equation}
	\begin{aligned}
		 \bm{R}_{\mathrm t}\bm{P}_{\mathrm t}\bm{\Gamma}_Q\bm{A}_{\mathrm{cp}}=\bm{P}_{\mathrm{t,cir}}\bm{\Gamma}_{Q,N} .
	\end{aligned} \label{eq:pulse_circularization}
\end{equation}
Consequently, the sensing reference waveform admits the second and more compact representation
\begin{equation}
	\begin{aligned}
		\boxed{\widetilde{\bm{x}}=\bm{P}_{\mathrm{t,cir}}\bm{\Gamma}_{Q,N}\bm{U}\bm{s} = \bm{p}_{\mathrm{t},QN} \circledast \bm{x}_{\mathrm{up}} 
			= \bm{p}_{\mathrm{t}} \circledast_{QN} \bm{x}_{\mathrm{up}} . }
	\end{aligned} \label{eq:x_tilde_circular}
\end{equation}
Here, $\circledast$ denotes circular convolution between two length-$QN$ vectors, whereas $\circledast_{QN}$ explicitly denotes $QN$-point circular convolution. In $\bm{p}_{\mathrm t}\circledast_{QN}\bm{x}_{\mathrm{up}}$, the finite-length pulse $\bm{p}_{\mathrm t}$ is zero-padded to length $QN$ before circular convolution.

Equation~\eqref{eq:x_tilde_circular} shows that the useful sensing reference waveform is equivalently obtained through $QN$-point circular pulse shaping of the upsampled useful signal block $\bm{x}_{\mathrm{up}}$. This equivalence is enabled by the CP inserted in the physical transmit chain. The resulting reference waveform $\widetilde{\bm{x}}$, rather than the complete transmitted waveform $\bm{x}_{\mathrm t}$, is used in the subsequent periodic sensing model and matched-filter processing.

\subsubsection{CP Removal}

The sensing receiver extracts from $\bm{y}_{\mathrm{s,full}}$ the same length-$QN$ absolute time interval used to construct $\widetilde{\bm{x}}$. Define
\begin{equation}
	\begin{aligned}
		\bm{R}_{\mathrm s}=\begin{bmatrix}\bm{0}_{QN\times QN_{\mathrm{cp}}}~\bm{I}_{QN}~\bm{0}_{QN\times(L_{\mathrm p}-1+\tau_{\max})}\end{bmatrix}\in\mathbb R^{QN\times K_{\mathrm s}}.
	\end{aligned} \label{eq:R_s}
\end{equation}
The useful sensing observation is
\begin{equation}
	\begin{aligned}
		\boxed{\bm{y}_{\mathrm s}=\bm{R}_{\mathrm s}\bm{y}_{\mathrm{s,lin}}\in\mathbb C^{QN\times1}.}
	\end{aligned} \label{eq:y_s_selected}
\end{equation}
The matrices $\bm{R}_{\mathrm t}$ and $\bm{R}_{\mathrm s}$ select the same absolute time interval, but operate on vectors of different lengths. Specifically, $\bm{x}_{\mathrm t}$ contains the pulse-shaping tail, whereas $\bm{y}_{\mathrm{s,lin}}$ additionally contains the delay-induced tail associated with $\tau_{\max}$.

Assume that the high-rate CP covers both the pulse-shaping memory and the maximum target delay, i.e.,
\begin{equation}
	\begin{aligned}
		QN_{\mathrm{cp}}\geq L_{\mathrm p}-1+\tau_{\max},
	\end{aligned} \nonumber \label{eq:sensing_cp_condition}
\end{equation}
then the physical aperiodic delay within the selected sensing interval is equivalent to a circular shift of the sensing reference waveform:
\begin{equation}
	\begin{aligned}
		\boxed{\bm{R}_{\mathrm s}\widetilde{\bm{J}}_{\tau_q}\bm{x}_{\mathrm t}=\bm{J}_{\tau_q}\widetilde{\bm{x}},}
	\end{aligned} \label{eq:sensing_circularization}
\end{equation}
where
\begin{equation}
	\begin{aligned}
		\bm{J}_{\tau_q}=\begin{bmatrix}\bm{0}_{\tau_q\times(QN-\tau_q)}&\bm{I}_{\tau_q}\\\bm{I}_{QN-\tau_q}&\bm{0}_{(QN-\tau_q)\times\tau_q}\end{bmatrix}\in\mathbb R^{QN\times QN}
	\end{aligned} \label{eq:J_circular}
\end{equation}
denotes the length-$QN$ circular delay matrix.

Since $\bm{x}_{\mathrm t}=\bm{P}_{\mathrm t}\bm{\Gamma}_Q\bm{A}_{\mathrm{cp}}\bm{U}\bm{s}$, the corresponding operator relation is
\begin{equation}
	\begin{aligned}
		 \bm{R}_{\mathrm s}\widetilde{\bm{J}}_{\tau_q}\bm{P}_{\mathrm t}\bm{\Gamma}_Q\bm{A}_{\mathrm{cp}}=\bm{J}_{\tau_q}\bm{R}_{\mathrm t}\bm{P}_{\mathrm t}\bm{\Gamma}_Q\bm{A}_{\mathrm{cp}} .
	\end{aligned} \nonumber \label{eq:sensing_operator_identity}
\end{equation}
Substituting \eqref{eq:sensing_circularization} into \eqref{eq:y_s_selected} gives
\begin{equation}
	\begin{aligned}
		\bm{y}_{\mathrm s}&=\sum_{q=1}^{K}\beta_q\bm{R}_{\mathrm s}\widetilde{\bm{J}}_{\tau_q}\bm{x}_{\mathrm t}+\bm{z}_{\mathrm s}\\&=\sum_{q=1}^{K}\beta_q\bm{J}_{\tau_q}\widetilde{\bm{x}}+\bm{z}_{\mathrm s},
	\end{aligned}
\end{equation}
where $\bm{z}_{\mathrm s}=\bm{R}_{\mathrm s}\bm{z}_{\mathrm{s,full}}$. Therefore, the useful sensing observation follows the equivalent periodic model
\begin{equation}
	\begin{aligned}
		\boxed{\bm{y}_{\mathrm s}=\sum_{q=1}^{K}\beta_q\bm{J}_{\tau_q}\widetilde{\bm{x}}+\bm{z}_{\mathrm s}.}
	\end{aligned} \label{eq:sensing_periodic_model}
\end{equation}
The physical propagation remains aperiodic and is described by $\widetilde{\bm{J}}_{\tau_q}$. The circular-delay structure in \eqref{eq:sensing_periodic_model} is induced jointly by the CP extension at the transmitter and the useful-window extraction at the sensing receiver.

\subsubsection{Periodic Matched Filtering}

Based on \eqref{eq:sensing_periodic_model}, the sensing receiver performs periodic matched filtering using the reference waveform $\widetilde{\bm{x}}$. For a candidate delay index $k\in\{0,1,\ldots,QN-1\}$, the corresponding matched-filter output is
\begin{equation}
	\begin{aligned}
		\widetilde{y}_k=\left(\bm{J}_k\widetilde{\bm{x}}\right)^{\mathrm H}\bm{y}_{\mathrm s}=\widetilde{\bm{x}}^{\mathrm H}\bm{J}_k^{\mathrm H}\bm{y}_{\mathrm s}.
	\end{aligned} \label{eq:periodic_MF}
\end{equation}
Define the periodic ACF of the reference waveform as
\begin{equation}
	\begin{aligned}
		R_{k}\triangleq\widetilde{\bm{x}}^{\mathrm H}\bm{J}_k^{\mathrm H}\widetilde{\bm{x}}.
	\end{aligned} \label{eq:periodic_ACF}
\end{equation}
Substituting \eqref{eq:sensing_periodic_model} into \eqref{eq:periodic_MF} yields
\begin{equation}
	\begin{aligned}
		\widetilde{y}_k=\sum_{q=1}^{K}\beta_q\widetilde{\bm{x}}^{\mathrm H}\bm{J}_k^{\mathrm H}\bm{J}_{\tau_q}\widetilde{\bm{x}}+\widetilde{z}_k,
	\end{aligned} \label{eq:MF_expanded}
\end{equation}
where $\widetilde{z}_k=\widetilde{\bm{x}}^{\mathrm H}\bm{J}_k^{\mathrm H}\bm{z}_{\mathrm s}$. Since
\begin{equation}
	\begin{aligned}
		\bm{J}_k^{\mathrm H}\bm{J}_{\tau_q}=\bm{J}_{\tau_q-k},
	\end{aligned}
\end{equation}
the matched-filter output becomes
\begin{equation}
	\begin{aligned}
		\boxed{\widetilde{y}_k=\sum_{q=1}^{K}\beta_qR_{ k-\tau_q}+\widetilde{z}_k.}
	\end{aligned} \label{eq:sensing_range_profile}
\end{equation}
Thus, the sensing range profile consists of delayed and scaled replicas of the periodic ACF of $\widetilde{\bm{x}}$. In particular, the $q$th target produces a peak around $k=\tau_q$, while the interference among targets is governed by the sidelobes of $R_k$.

The periodic matched-filter outputs over all candidate delays can be efficiently evaluated through the FFT as
\begin{equation}
	\begin{aligned}
	 \widetilde{\bm{y}}=\operatorname{IFFT}\left(\operatorname{FFT}(\bm{y}_{\mathrm s})\odot\operatorname{FFT}(\widetilde{\bm{x}})^{*}\right) ,
	\end{aligned} \label{eq:sensing_fft}
\end{equation}
where $\widetilde{\bm{y}}=[\widetilde{y}_0,\widetilde{y}_1,\ldots,\widetilde{y}_{QN-1}]^{\mathrm T}$.

\subsubsection{Range Estimation}

The squared magnitude of the periodic matched-filter output defines the range profile:
\begin{equation}
	\begin{aligned}
		P_{k}=|\widetilde{y}_k|^2,\qquad k=0,1,\ldots,QN-1.
	\end{aligned}
	\label{eq:range_profile_power}
\end{equation}
For a single-target scenario, the target delay is estimated by searching for the maximum of the range profile:
\begin{equation}
	\begin{aligned}
		\widehat{\tau}=\mathop{\arg\max}\limits_kP_{k}.
	\end{aligned}
	\label{eq:delay_estimation}
\end{equation}
The corresponding range estimate is
\begin{equation}
	\begin{aligned}
		\widehat{R}=\frac{c_0T_{\mathrm s}}{2}\widehat{\tau}=\frac{c_0}{2F_{\mathrm s}}\widehat{\tau}.
	\end{aligned}
	\label{eq:range_estimation}
\end{equation}
For a multiple-target scenario, the target delays are obtained by detecting the local peaks of $P_k$ and converting their delay indices into physical ranges according to \eqref{eq:range_delay_mapping}.

In summary, the two receivers process different observations of this common transmitted waveform. The communication signal propagates through a frequency-selective multipath channel. The communication receiver then applies the receive matched filter, compensates for the overall timing offset, downsamples the filtered signal to the symbol rate, removes the CP, and performs channel equalization and symbol detection. The cascade of transmit pulse shaping, physical multipath propagation, receive matched filtering, timing alignment, and downsampling forms a symbol-rate equivalent linear channel. When the CP is sufficiently long, this equivalent channel becomes circulant after CP removal, thereby enabling low-complexity frequency-domain equalization for OFDM signaling.

The sensing receiver observes delayed and scaled replicas of the same transmitted waveform reflected by multiple targets. Unlike the communication receiver, it does not use the receive pulse matched filter to recover the communication symbols. Instead, it extracts a high-rate useful observation block and correlates this block with a locally reconstructed sensing reference waveform. The sensing reference is obtained from the actual transmitted waveform by selecting the useful interval after the high-rate CP. When the CP covers the pulse-shaping memory, this reference is equivalently generated by circular pulse shaping of the upsampled communication signal with CP. Moreover, when the CP also covers the maximum target delay, the physical aperiodic target delays become circular shifts of the sensing reference waveform within the selected observation interval after removing CP. Periodic matched filtering can then be employed to construct the range profile and estimate the target delays.

\section{End-to-End Equivalent Model}
This section consolidates the preceding transmitter, communication receiver, and sensing receiver models into end-to-end input--output representations. Both links originate from the same communication-centric ISAC transmitter, whereas their receivers perform different signal-processing operations according to their respective objectives.

\subsection{End-to-End Communication Model}
For the communication link, the transmitted waveform successively undergoes the frequency-selective communication channel, receive matched filtering, timing compensation, downsampling, and CP removal. Combining these operations gives the complete end-to-end communication model as
\begin{equation}
	\begin{aligned}
		\bm{y}&=\overline{\bm{R}}_{\mathrm{cp}}\bm{D}_{Q,n_0}\bm{P}_{\mathrm r}\bm{H}\bm{P}_{\mathrm t}\bm{\Gamma}_{Q}\bm{A}_{\mathrm{cp}}\bm{U}\bm{s}+\bm{w}_{y}\\
		&=\overline{\bm{R}}_{\mathrm{cp}}\bm{H}_{\mathrm{eq,lin}}\bm{A}_{\mathrm{cp}}\bm{U}\bm{s}+\bm{w}_{y}\\
		&=\bm{H}_{\mathrm{eq,cir}}\bm{U}\bm{s}+\bm{w}_{y},
	\end{aligned} \label{eq:communication_end_to_end}
\end{equation}
where the second equality follows from the definition of the symbol-rate equivalent linear channel $\bm{H}_{\mathrm{eq,lin}}$, while the last equality follows from the CP-induced circularization relation $\overline{\bm{R}}_{\mathrm{cp}}\bm{H}_{\mathrm{eq,lin}}\bm{A}_{\mathrm{cp}}=\bm{H}_{\mathrm{eq,cir}}$. Therefore, the physical cascade consisting of upsampling, transmit pulse shaping, the communication channel, receive matched filtering, timing compensation, and downsampling is represented at the symbol rate by the circulant matrix $\bm{H}_{\mathrm{eq,cir}}$. Equation \eqref{eq:communication_end_to_end} is the complete signal-processing interpretation of \eqref{eq:comm_useful_block}.

For OFDM signaling, $\bm{U}=\bm{F}_{N}^{\mathrm H}$. Since $\bm{H}_{\mathrm{eq,cir}}$ is circulant, it can be diagonalized by the normalized DFT matrix as
\begin{equation}
	\begin{aligned}
		\bm{H}_{\mathrm{eq,cir}}=\bm{F}_{N}^{\mathrm H}\bm{\Lambda}_{\mathrm{eq}}\bm{F}_{N},
	\end{aligned}
	\label{eq:communication_channel_diagonalization}
\end{equation}
where $\bm{\Lambda}_{\mathrm{eq}}$ is a diagonal matrix containing the frequency response of the equivalent channel. Applying the $N$-point DFT to \eqref{eq:communication_end_to_end} yields
\begin{equation}
	\begin{aligned}
		\bm{Y}=\bm{F}_{N}\bm{y}=\bm{\Lambda}_{\mathrm{eq}}\bm{s}+\bm{W}_{y},
	\end{aligned}
	\label{eq:communication_frequency_domain}
\end{equation}
where $\bm{W}_{y}=\bm{F}_{N}\bm{w}_{y}$. Consequently, the frequency-selective channel is decomposed into $N$ parallel scalar subchannels, and the transmitted symbols can be recovered using one-tap frequency-domain equalization.

\subsection{End-to-End Sensing Model}
Unlike the communication receiver, the sensing receiver does not attempt to recover the information symbols. Instead, it exploits the known sensing reference waveform $\widetilde{\bm{x}}$ to estimate the target delays through periodic matched filtering. As established in \eqref{eq:x_tilde_circular}, the sensing reference waveform is determined by
\begin{equation}
	\begin{aligned}
		\boxed{\widetilde{\bm{x}}=\bm{P}_{\mathrm{t,cir}}\bm{\Gamma}_{Q,N}\bm{U}\bm{s}=\bm{p}_{\mathrm{t},QN}\circledast\bm{x}_{\mathrm{up}}=\bm{p}_{\mathrm{t}}\circledast_{QN}\bm{x}_{\mathrm{up}}.}
	\end{aligned}  \nonumber
\end{equation}
This equation provides the end-to-end mapping from the communication symbols $\bm{s}$ to the high-rate sensing reference waveform $\widetilde{\bm{x}}$. It incorporates the modulation basis, upsampling, and transmit pulse shaping within the useful sensing interval.

After the sensing receiver selects the useful high-rate observation interval, the physical target delays covered by the CP are converted into circular shifts of $\widetilde{\bm{x}}$. Consequently, the received sensing signal is expressed as
\begin{equation}
	\begin{aligned}
		\bm{y}_{\mathrm s}=\sum_{q=1}^{K}\beta_q\bm{J}_{\tau_q}\widetilde{\bm{x}}+\bm{z}_{\mathrm s}.
	\end{aligned}  \nonumber
\end{equation}
Therefore, the $q$th target contributes a circularly shifted replica of the common sensing reference waveform, with delay index $\tau_q$ and complex reflection coefficient $\beta_q$.

The periodic ACF of the sensing reference waveform at delay index $k$ is defined as
\begin{equation}
	\begin{aligned}
		R_k=\widetilde{\bm{x}}^{\mathrm H}\bm{J}_{k}^{\mathrm H}\widetilde{\bm{x}}.
	\end{aligned} \nonumber
\end{equation}
Applying periodic matched filtering to $\bm{y}_{\mathrm s}$ gives the sensing range profile
\begin{equation}
	\begin{aligned}
		\widetilde y_k=\sum_{q=1}^{K}\beta_qR_{k-\tau_q}+\widetilde z_k.
	\end{aligned} \nonumber
\end{equation}
Hence, each target produces a scaled and shifted copy of the periodic ACF $R_k$. The locations of the resulting peaks determine the target delays, whereas the sidelobe structure of $R_k$ governs the mutual interference among targets with different reflection strengths.

\begin{figure}[t]
	\centering
	\subfigure[16PSK.]{
	\begin{minipage}[t]{0.9\linewidth}
		\centering
		\includegraphics[width=1\textwidth]{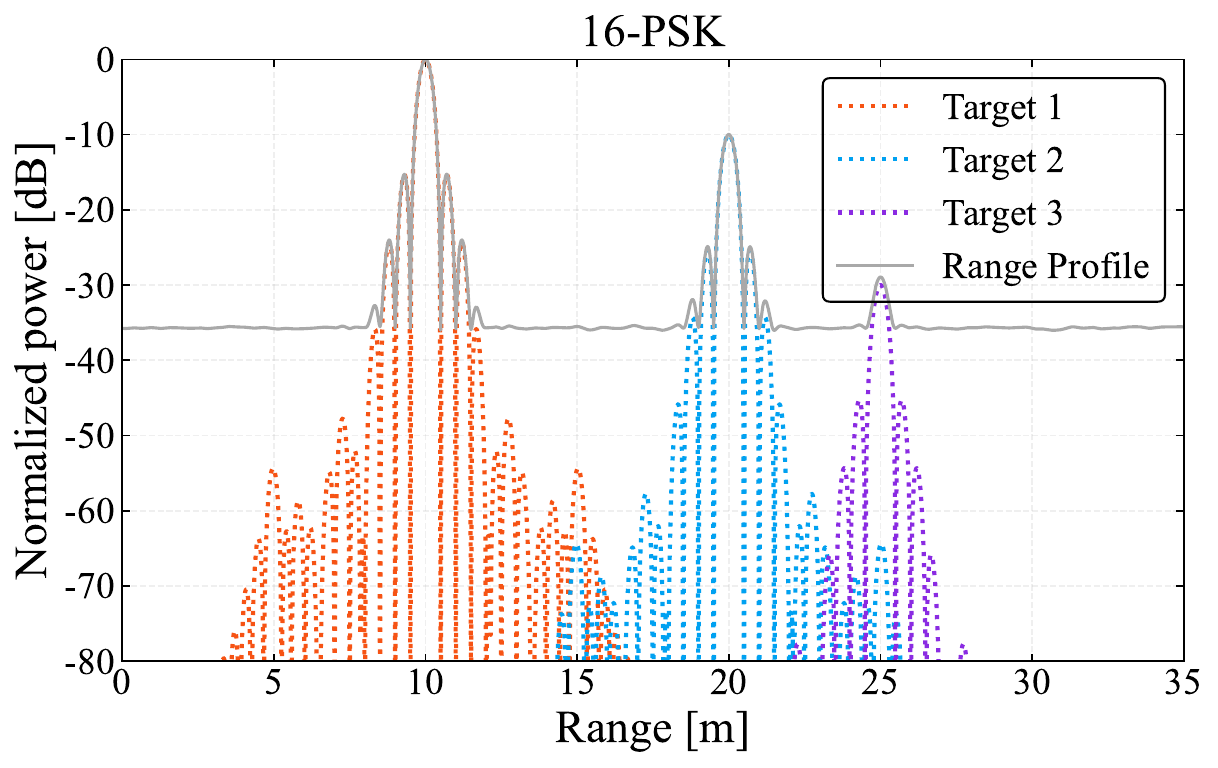} 
	\end{minipage}
	}
	\subfigure[4QAM.]{
		\begin{minipage}[t]{0.9\linewidth}
			\centering
			\includegraphics[width=1\textwidth]{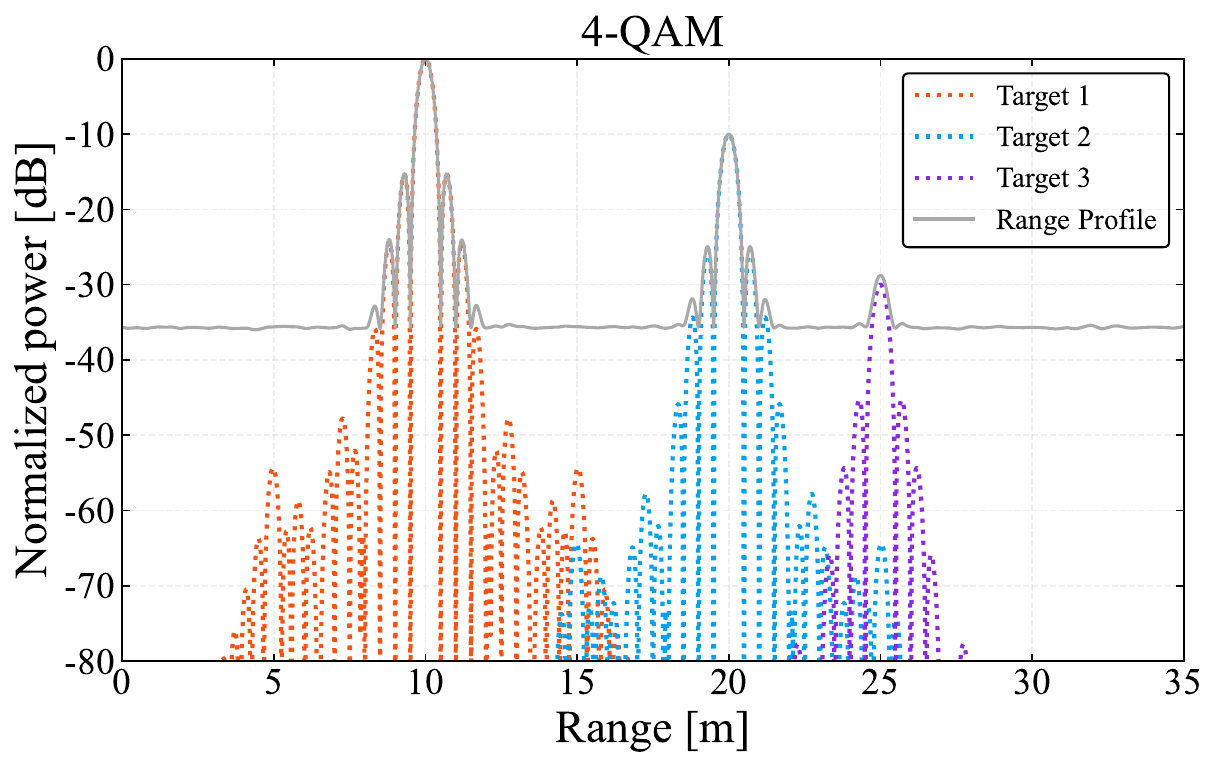} 
		\end{minipage}
	}
	\subfigure[16QAM.]{
		\begin{minipage}[t]{0.9\linewidth}
			\centering
			\includegraphics[width=1\textwidth]{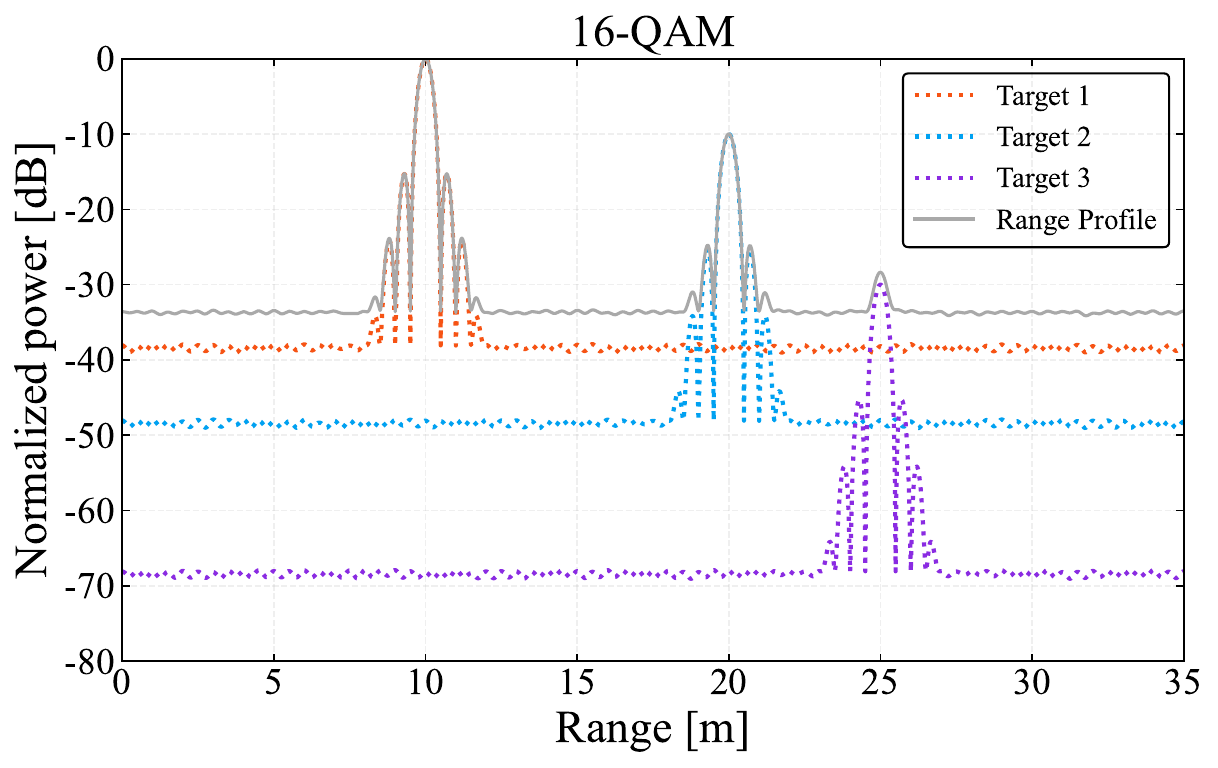}
		\end{minipage}
	}
	\centering
	\caption{An example of the range profile for OFDM signal with PSK/QAM constellation, including 3 targets located at 10m, 20m, and 25m.}
	\label{fig:ACF}
\end{figure}

\begin{figure}[t]
	\centering
	\subfigure[SER of PSK.]{
		\begin{minipage}[t]{0.9\linewidth}
			\centering
			\includegraphics[width=1\textwidth]{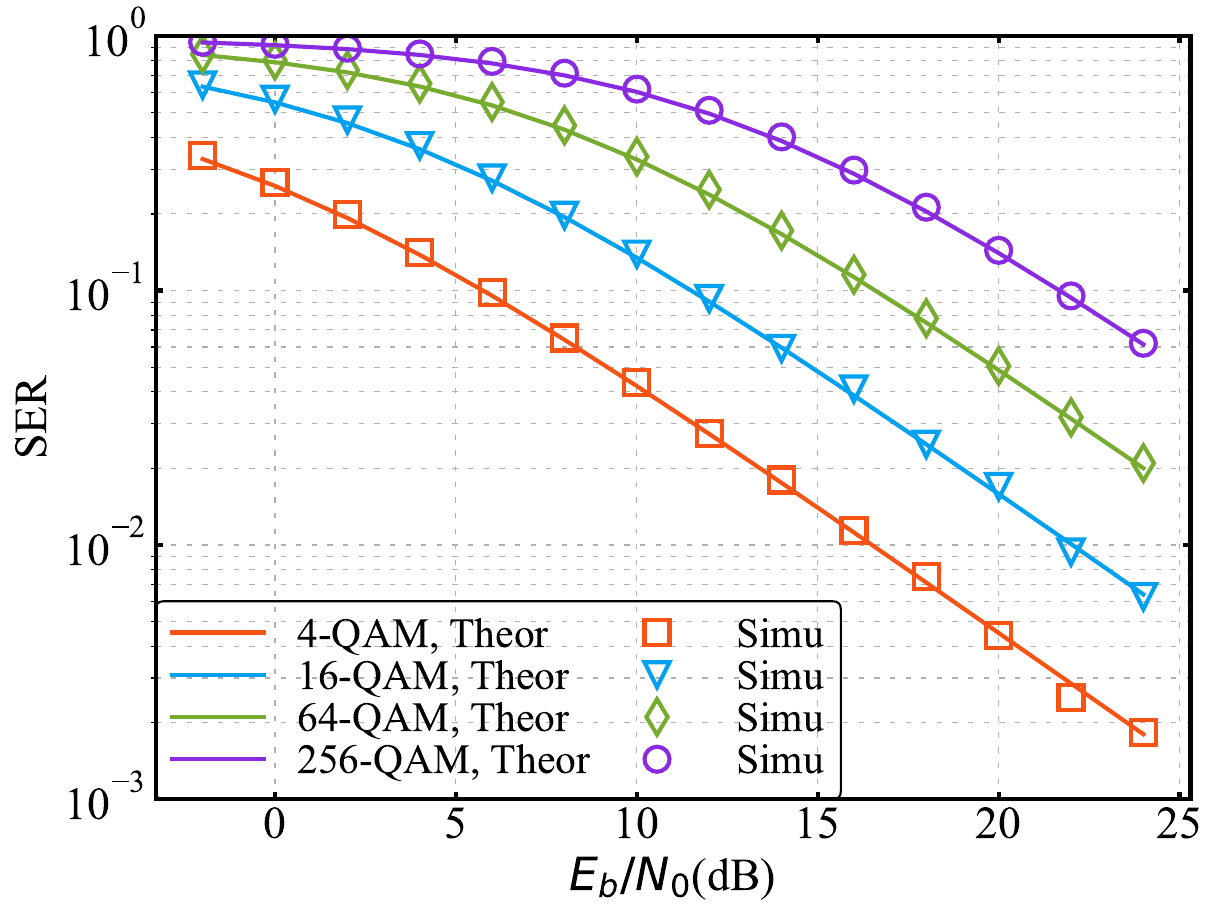} 
		\end{minipage}
	}
	\subfigure[SER of QAM.]{
		\begin{minipage}[t]{0.9\linewidth}
			\centering
			\includegraphics[width=1\textwidth]{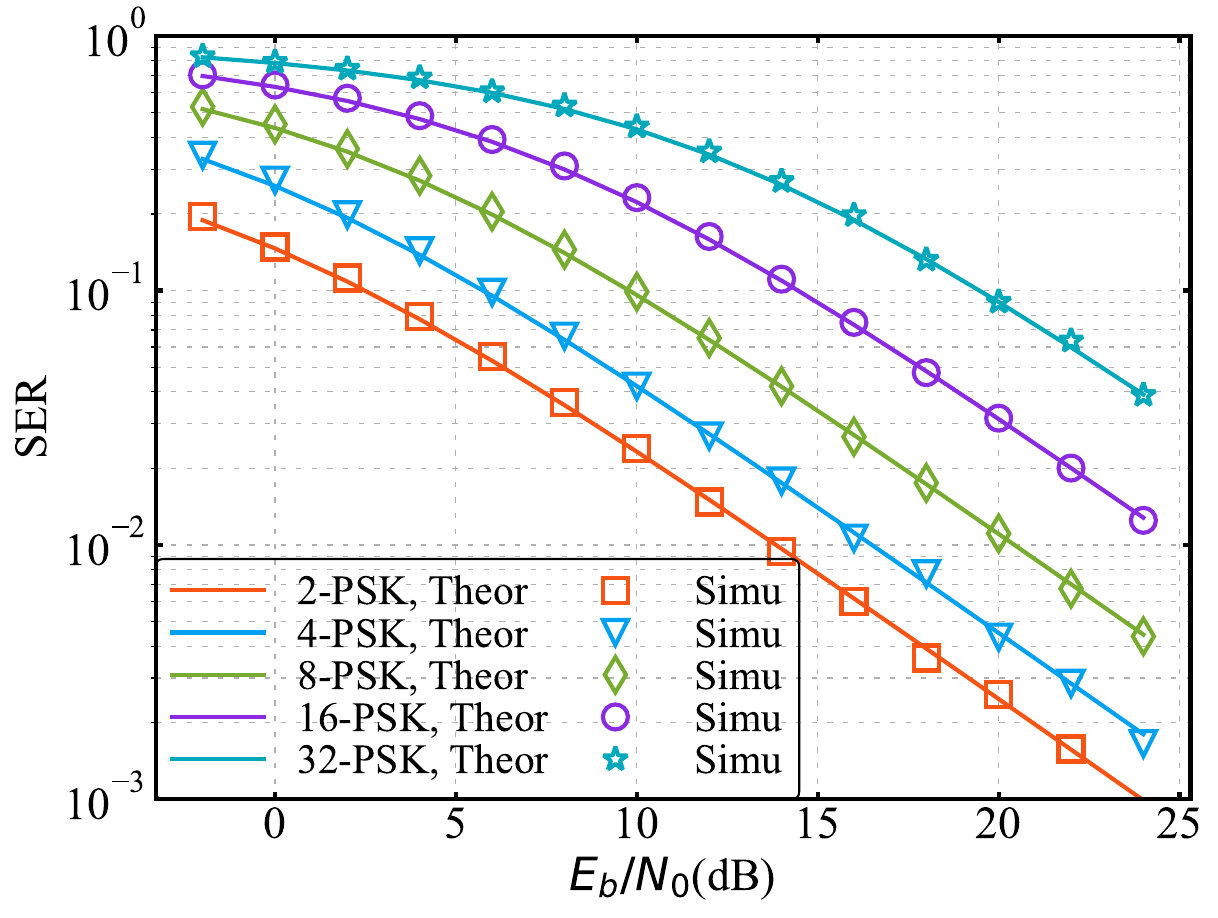}
		\end{minipage}
	}
	\centering
	\caption{SER performance of the complete communication link over a frequency-selective Rayleigh fading channel.}
	\label{fig:SER_QAM}
\end{figure}

\section{Simulation Results}
In this section, numerical simulations are presented to validate the sensing and communication models developed in the preceding sections. The sensing experiment verifies the CP-supported periodic target-echo model and the corresponding range-profile formulation, whereas the communication experiment verifies the complete CP-OFDM transmission chain with Nyquist pulse shaping over a frequency-selective Rayleigh fading channel.

Both experiments employ the same transmitter configuration, with $N=2048$ subcarriers, an oversampling factor of $Q=20$, and a CP length of $N_{\mathrm{cp}}=80$. The transmit pulse is a unit-energy root-raised-cosine~(RRC) pulse with a roll-off factor of $0.35$ and a truncation span of $20$ symbol periods, yielding $L_{\mathrm p}=401$ samples. The communication receiver employs the corresponding pulse matched filter, with $L_{\mathrm r}=401$, and downsamples its output with sampling phase $n_0=0$. The physical communication channel contains $L=5$ consecutive high-rate taps. Consequently, the combined impulse response has length $L_c=805$, and the symbol-rate equivalent channel has length $L_{\mathrm{eq}}=41$. The selected CP therefore satisfies the communication requirement $N_{\mathrm{cp}}\geq L_{\mathrm{eq}}-1$.

The high-rate sampling frequency is $F_{\mathrm s}\approx5.996$~GHz, corresponding to a range-grid spacing of $\Delta R=0.025$~m. The sensing reference waveform contains $QN=40960$ samples. For the target configuration considered below, the maximum delay index is $\tau_{\max}=1000$, and the sensing CP requirement is also satisfied, since $QN_{\mathrm{cp}}=1600\geq L_{\mathrm p}-1+\tau_{\max}=1400$. The sensing results are averaged over $2000$ independent realizations of the communication data and receiver noise, whereas the communication SER is evaluated using $2000$ OFDM blocks for each constellation and each $E_b/N_0$ value.

\subsection{Sensing Performance}
Figs.~\ref{fig:ACF}(a), \ref{fig:ACF}(b), and \ref{fig:ACF}(c) present the range profiles obtained using 16-PSK, 4-QAM, and 16-QAM constellations, respectively. In each experiment, three targets are located at $10$~m, $20$~m, and $25$~m, with reflection amplitudes of $0$~dB, $-10$~dB, and $-30$~dB and zero reflection phases. Their corresponding high-rate delay indices are $\tau_q=400$, $800$, and $1000$, respectively. Doppler shifts are not considered. The sensing input SNR is set to $-10$~dB, measured using the average power of the total noiseless echo within the useful observation interval. The plotted profiles are obtained by averaging the squared magnitudes of the periodic matched-filter outputs. Within each subfigure, all curves are normalized by the maximum of the average total range profile.

The individual target responses in Fig.~\ref{fig:ACF} are determined by shifted replicas of the periodic ACF of the sensing reference waveform $\widetilde{\bm{x}}$. The complex responses add coherently before the squared magnitude is evaluated to form the total range profile. The individual response peaks occur at the prescribed target locations, which verifies the delay-to-range mapping and the periodic target-echo model in \eqref{eq:sensing_periodic_model}. In particular, the result confirms that, after selecting the useful sensing interval, a physical linear delay covered by the CP is equivalently represented by a circular shift of $\widetilde{\bm{x}}$.

The numerical result also agrees with the range-profile expression in \eqref{eq:range_profile_power}. Specifically, the contribution of the $q$th target to the complex matched-filter output is determined by $\beta_qR_{k-\tau_q}$, such that its peak location is specified by $\tau_q$, its amplitude is scaled by $\beta_q$, and its interference with the other targets is governed by the sidelobes of $R_k$. Therefore, the ability to resolve a weak target in the vicinity of a strong target depends directly on the periodic ACF of the actual sensing reference waveform.

Comparing the three subfigures, 16-PSK and 4-QAM exhibit similar range profiles, whereas 16-QAM introduces a higher sidelobe floor in the individual target responses. This behavior is consistent with the iceberg analysis in~\cite{liu2025uncovering}: the constant-modulus 16-PSK and 4-QAM constellations have a normalized fourth moment of $1$, whereas that of 16-QAM is $1.32$. The amplitude fluctuations of 16-QAM therefore introduce an additional random sidelobe component. As a result, the weak target at $25$~m is more susceptible to masking by the sidelobes of stronger targets than in the 4-QAM and 16-PSK cases. This observation is consistent with the deterministic-random tradeoff discussed in~\cite{liu2025cpofdm}, illustrating the sensing benefit of suppressing data-induced fluctuations in the waveform autocorrelation.

\subsection{Communication Performance Over Rayleigh Fading Channels}
Fig.~\ref{fig:SER_QAM} shows the symbol error rate~(SER) performance of the complete communication link over a frequency-selective Rayleigh fading channel. The considered transmission process includes OFDM modulation, CP insertion, upsampling, transmit pulse shaping, physical multipath propagation, receive matched filtering, timing compensation, downsampling, CP removal, DFT processing, and one-tap frequency-domain equalization. The channel coefficients are independently generated according to a circularly symmetric complex Gaussian distribution and remain fixed within each OFDM block. A deterministic ensemble normalization is applied such that $\mathbb{E}[\|\bm{h}_{\mathrm{eq}}\|_2^2]=1$, and perfect equivalent-channel knowledge is assumed at the communication receiver.

In Fig.~\ref{fig:SER_QAM}(a), the simulated SER curves are compared with the corresponding theoretical Rayleigh fading results for 4-,~16-,~64-,~and 256-QAM. The theoretical curves account for the CP overhead and the average equivalent-channel gain and filtered-noise variance on each subcarrier, followed by averaging over all subcarriers. The close agreement between the simulated and theoretical curves validates both the high-rate physical implementation and the symbol-rate equivalent model in \eqref{eq:communication_end_to_end}. In particular, it confirms that the cascade of pulse shaping, the frequency-selective channel, receive matched filtering, timing compensation, and downsampling can be represented by $\bm{H}_{\mathrm{eq,lin}}$, and that a sufficiently long CP converts this matrix into the circulant equivalent channel $\bm{H}_{\mathrm{eq,cir}}$ over the useful received block.
For the considered QAM constellations, higher modulation orders exhibit higher SERs at a given $E_b/N_0$, while all curves decrease as $E_b/N_0$ increases.
Fig.~\ref{fig:SER_QAM}(b) presents the corresponding results for 2-, 4-, 8-, 16-, and 32-PSK, for which close agreement between the simulated and theoretical SER curves is also observed.  

\section{Conclusion}
This paper developed a DSP-oriented discrete-time transceiver framework for single-antenna communication-centric ISAC. Starting from a common communication transmitter, the complete waveform generation process was described through modulation, CP insertion, upsampling, and finite-length linear pulse shaping. At the communication receiver, transmit pulse shaping, physical multipath propagation, receive matched filtering, timing alignment, and downsampling were combined into a symbol-rate equivalent linear channel. Under a sufficient CP-length condition, CP insertion and removal convert this equivalent linear channel into a circulant channel, enabling frequency-domain equalization for OFDM signaling. At the sensing receiver, the high-rate sensing reference waveform was extracted from the actual transmitted waveform and equivalently represented by circular pulse shaping of the upsampled useful signal block. It was further shown that the physical aperiodic target delays covered by the CP become circular shifts of the sensing reference waveform over the useful observation interval, leading to the periodic matched-filtering model and the corresponding range profile. Numerical results validated the sensing model through target range estimation and verified the communication model through the agreement between the simulated and theoretical SER performance of pulse-shaped CP-OFDM over frequency-selective Rayleigh fading channels. The developed framework therefore provides the DSP foundation connecting physical pulse-shaped transceiver operations to the periodic signal models used for random ISAC waveform analysis.

\bibliographystyle{ieeetran}
\bibliography{ref_abbreviation}	

\begin{IEEEbiography}[{\includegraphics[width=1in,height=1.25in,clip,keepaspectratio]{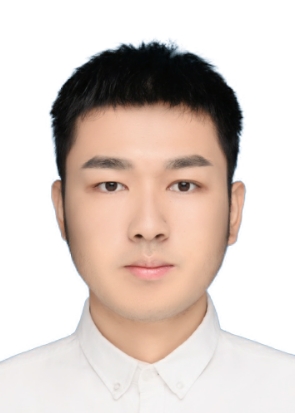}}]{Junjie~Chen} received the B.Eng. degree from the Beijing Institute of Technology (BIT), Beijing, China, in 2017, and the M.E. degree from the University of Science and Technology of China (USTC), Hefei, China, in 2020. 
From 2020 to 2022, he was with the China Electronics Technology Group Corporation (CETC), where he conducted research on radar and communication systems. 
He received the Ph.D. degree from Sun Yat-sen University (SYSU), Guangzhou, China, in 2026. 
He is currently a lecturer with the School of Computing and Artificial Intelligence, Jiangxi University of Finance and Economics, Nanchang, China. 
His research interests include signal processing, wireless communications, and machine learning. \end{IEEEbiography}

\end{document}